# A comparison between behavioral similarity methods vs standard deviation method in predicting time series dataset, case study of finance market

**Mahdi Goldani**

m.goldani@hsu.ac.ir

**Abstract:**

In statistical modeling, prediction and explanation are two fundamental objectives. When the primary goal is forecasting, it is important to account for the inherent uncertainty associated with estimating unknown outcomes. Traditionally, confidence intervals constructed using standard deviations have served as a formal means to quantify this uncertainty and evaluate the closeness of predicted values to their true counterparts. This approach reflects an implicit aim to capture the behavioral similarity between observed and estimated values. However, advances in similarity-based approaches present promising alternatives to conventional variance-based techniques, particularly in contexts characterized by large datasets or a high number of explanatory variables. This study aims to investigate which methods—either traditional or similarity-based—are capable of producing narrower confidence intervals under comparable conditions, thereby offering more precise and informative intervals. The dataset utilized in this study consists of U.S. mega-cap companies, comprising 42 firms. Due to the high number of features, interdependencies among predictors are common; therefore, Ridge Regression is applied to address this issue. The research findings indicate that the σ-based method and LCSS exhibit the highest coverage among the analyzed methods, although they produce broader intervals. Conversely, DTW, Hausdorff, and TWED deliver narrower intervals, positioning them as the most accurate methods, despite their medium coverage rates. Ultimately, the trade-off between interval width and coverage underscores the necessity for context-aware decision-making when selecting similarity-based methods for confidence interval estimation in time series analysis.

**Keywords:** statistical modeling, prediction, uncertainty, confidence interval

1. **Introduction**

Everyday scenarios involve uncertainty across diverse fields—from investment decisions and medical diagnoses to sports outcomes and weather forecasts. In all cases, the goal is to make informed decisions using available data and domain knowledge despite inherent uncertainties (Abdar et al, 2021). Researchers have proposed various models, including regression, machine learning, and neural networks, to achieve accurate predictions. However, even the most advanced methods cannot predict exact values with certainty; inherent uncertainty always exists. Therefore, since predictions made by such models are subject to noise and inference error, uncertainty quantification (UQ) becomes essential (Malinin, 2019). To quantify this prediction uncertainty, metrics such as Mean Squared Error (MSE) and Mean Absolute Percentage Error (MAPE) are commonly used, measuring the error between predictions and actual outcomes (Quan et al, 2014). Prediction intervals (PIs) are among the most widely used techniques for uncertainty quantification (UQ) in the literature (Kabir et al, 2018).

Prediction intervals (PIs) have been extensively used to quantify that uncertainty for over 50 years (Maskey et al, 2004). A prediction interval provides a range expected to contain future observations, thereby quantifying prediction uncertainty. This approach overcomes point prediction limits and is now vital in economics, food science, tourism, healthcare, energy, and even compression algorithms (Zhang et al, 2021). Despite their significance, the evaluation and optimization of prediction intervals remain areas ripe for exploration, particularly through innovative methodologies like distance-based methods.

Optimal confidence intervals are crucial for accurately estimating parameters across various fields, including reliability engineering, diagnostics, and economic forecasting. Optimal confidence intervals (CIs) achieve high coverage probability (e.g., 95%) with minimal width. Wider intervals indicate greater uncertainty about the estimated parameter (Hazra, 2017). The bootstrap weighted-norm method is a robust technique for accurately estimating confidence intervals in parameter estimation (Cui and Xie, 2023). In diagnostic measures, both delta method and parametric bootstraps are suitable for post-hoc calculation of confidence intervals, depending on the sample size, the distribution of marker values, and the correctness of model assumptions (Thiele & Hirschfeld, 2023). Furthermore, a moderate deviation principle-based approach provides statistically optimal confidence intervals for non-parametric estimations, ensuring minimal mischaracterization and robust performance across diverse models (Ganguly & Sutter, 2023). In agricultural price forecasting, a trade-off model between accuracy and informativeness enables the evaluation of optimal confidence intervals and outperforms traditional approaches such as the equal probability and shortest interval methods (Wang et al., 2016). Collectively, these studies highlight the importance of tailored methodologies for constructing reliable confidence intervals across different applications. Exact intervals are the optimal methods for evaluating confidence intervals in hypergeometric settings, as they ensure proper coverage and offer a complete and admissible solution set for parameter inference (Inthout et al, 2016). Khosravi et al. (2011) examine four major methods for constructing prediction intervals (PIs) in neural network forecasting including delta, Bayesian, bootstrap, and mean-variance estimation (MVE) across 12 synthetic and real-world datasets. Results show that the delta and Bayesian methods produce the highest-quality and most consistent PIs, while the MVE and bootstrap methods offer advantages in terms of lower computational cost and better adaptability to data variability.

Conventional confidence interval needs statistical inference or error distribution assumptions, unlike these methods Wan et al. (2014) propose RAMODO, a framework that integrates representation learning with outlier detection using random distance-based methods. The findings demonstrate that the hybrid ELM-PSO approach provides a robust and adaptable solution for probabilistic wind power forecasting, showing significant promise for real-world implementation in contemporary power systems. Moreover, Quan et al. (2014) The introduced method, known as Lower Upper Bound Estimation (LUBE), is applied and enhanced to generate prediction intervals using neural network models. Validated with real-world data from Singapore, New South Wales, and Capital Wind Farm, the PSO-based LUBE approach demonstrates the ability to produce high-quality prediction intervals for both load and wind power forecasting, delivering improved accuracy with low computational overhead.

Traditionally, confidence intervals have been constructed using standard deviation, which quantifies the deviation of estimated values from actual outcomes—essentially measuring how closely the predicted behavior aligns with the true behavior of a variable. However, when certain conditions are met—such as a sufficiently large dataset or a high number of behavior-influencing variables—similarity-based methods can be applied using the same underlying logic as standard deviation. This research aims to determine which prediction methods are capable of producing narrower confidence intervals while maintaining constant type I and type II error rates.

1. Database

## 2-1. Database

This study is grounded in the U.S. finance market, spanning from April 7, 2020, to April 4, 2025. The closing prices of companies classified under the Mega Market Capitalization category were utilized. Apple's closing price was selected as the target value due to its stock performance often reflecting broader market trends, thereby making it a strong representative of market sentiment. The data were extracted from the NASDAQ database. Using the Stock Screener tool on the website, all active U.S. stock market companies within the Mega Cap category were filtered, resulting in a selection of 42 companies. The data spans the last five years with daily frequency, allowing for highly precise analysis of market fluctuations and enabling accurate statistical analyses and algorithmic applications.

## 2-2. Prediction model

Forecasting in finance involves a high number of variables, such as macroeconomic data, microeconomic data, earnings reports, and technical indicators. Multicollinearity and dependencies among predictors are common in financial datasets, which makes the use of multivariate regression models like OLS less appropriate or reliable (Chan et al, 2022). From a technical standpoint in linear regression with multicollinearity issue, when the matrix $X^TX$ is singular, the standard Ordinary Least Squares (OLS) estimators cannot be applied because they require the inverse of $X^TX$. in this case the ridge version for estimators is more stable and can overcome this problem (Arashi et al, 2021). Multicollinearity can cause issues for each coefficient, including inaccurate estimates, excessive growth of standard deviations, incorrect t-tests, and inaccurate confidence intervals (Mermi et al, 2024).

Ridge estimation is a regularization technique aimed at stabilizing parameter estimates by shrinking them or their linear combinations. This method provides more reliable estimators with reduced variance compared to ordinary least squares, especially in the presence of multicollinearity. By adding the k (where k>0) to the $X^TX$ matrix, one can control the amount of shrinkage of the regression coefficients:

$$\widehat{\beta_k} = (X^TX + kI)\hat{X}Y \qquad (1)$$

*"k" is* the shrinkage parameter, which controls the amount of shrinkage of the regression coefficients, and "**I**" is the identity matrix (Cule and De Iorio, 2013). As seen in Equation (1), the ridge regression approach involves adding a small positive number (k) to the diagonal elements of the ($X^TX$) matrix. This prevents the variance of the regression coefficients from overinflating due to multicollinearity.

## 2-3. Confidence intervals

Predictions are inherently associated with uncertainty. Unlike point forecasts, prediction intervals (PIs) serve as a powerful tool for modeling uncertainty. By definition, a PI consists of lower and upper bounds that bracket an unknown future value with a specified probability — typically expressed as a confidence level of (1−α) % (Quan et al, 2014). A desirable prediction method achieves narrower intervals while controlling Type I (false positive) and Type II (false negative) error probabilities.

Traditional form of confidence intervals uses standard deviation to evaluate CI. The margin of error (ME) in a confidence interval is determined using the Z-value, the standard deviation (SD) of the sample, and the sample size (N), and is given by the formula:

$$ME = Z \times \frac{SD}{\sqrt{N}} \tag{2}$$

The lower and upper bounds of a prediction interval are obtained by subtracting and adding the margin of error (ME) to the predicted value, respectively.

If $\hat{y}$ is the predicted value:

$lower\ bound = \hat{y} - ME$

$upper\ bound = \hat{y} + ME \tag{3}$

So, the prediction interval is:

$$PI = (\hat{y} - ME, \hat{y} + ME) \tag{4}$$

The Z-value is determined by the chosen confidence level. It is important to note that confidence intervals are statistically accurate only when sampling from a normally distributed population. For non-normally distributed populations, they become approximately valid when the sample size is sufficiently large (Simundic, 2008).

## 2-4. Distance-based methods

The construction of optimal confidence intervals (minimal width with correct coverage probability) as a statistical method relies critically on assumptions: Normally distributed and sufficiently large samples. Violating these assumptions compromises either coverage accuracy or interval efficiency. This article introduces distribution-free methods for confidence interval estimation that require no parametric assumptions (e.g., normality, large-sample approximations), leveraging distance-based metrics to achieve statistically valid inference.

Based on the same principles underlying the standard method for calculating confidence intervals, distance-based methods offer an alternative approach for constructing confidence intervals by measuring the distance between predicted values and observed data. Time series similarity measurement serves as the foundation for clustering and classification tasks by quantifying distances between temporal sequences. This metric plays a critical role in temporal pattern analysis, where it functions as a fundamental measure for statistical inference about cross-dataset relationships. Recent proliferation of data collection has significantly expanded time series availability, driving demand for analytical tasks like regression, classification, clustering, and segmentation. These applications universally require specialized distance metrics to quantify inter-series similarity, making methodological research in this area essential (Goldani and Asadi, 2024).

Current similarity measures fall into three primary categories:

1. Step-by-step methods (e.g., pointwise comparisons)
2. Distribution-based approaches (statistical property matching)
3. Geometric techniques (shape/trajectory alignment)

**2-4-1. Stepwise Metrics**

These metrics compare time-series samples one by one based on their time indices [20]. A significant limitation of these methods is the requirement for identical sample sizes in the time series. The most notable stepwise metrics are Euclidean Distance and Correlation Coefficient, which are detailed below.

- o The Euclidean Distance is the simplest measure for comparing time series. It calculates the shortest distance between two points in Euclidean space using the Pythagorean theorem. The Euclidean Distance between two time series x and y of length n is defined as:

$$Deuc = (\sum_{i=1}^{n}(x_i - y_i))^{1/2} \tag{5}$$

This distance is widely used due to its simplicity and ease of understanding. However, a key limitation of Euclidean Distance is its sensitivity to time-axis transformations, such as scaling and shifting [21]. Moreover, it cannot compare time series with different sample sizes. As it relies on point-to-point mapping, it is highly sensitive to noise and temporal misalignments, thus making it unsuitable for handling local shifts in time.
A straightforward extension of Euclidean Distance is to calculate the similarity using extracted features rather than raw time-series data.

**2-4-2. Elastic metrics**

These metrics adjust the time axis by stretching or compressing it to minimize the effect of local variations. These methods are particularly effective for handling non-linear distortions on time. The most notable elastic methods include Dynamic Time Warping (DTW), Longest Common Subsequence (LCSS), and others.

- o Dynamic Time Warping (DTW) is an algorithm for measuring similarity between time series that may vary in speed or timing. Unlike Euclidean Distance, DTW aligns sequences non-linearly by stretching or compressing the time axis to find the optimal alignment. The cumulative distance is calculated as:

$$DISTMATRIX = \begin{bmatrix} d(x_1,y_1) & d(x_1,y_2) & \ldots & d(x_1,y_m) \\ d(x_2,y_1) & d(x_2,y_2) & \ldots & d(x_2,y_m) \\ d(x_n,y_1) & d(x_n,y_2) & \ldots & d(x_n,y_m) \end{bmatrix} \tag{6}$$

$$\begin{cases} r(i,j) = d(i,j) + \min\{r(i-1,j), r(i,j-1), r(i-1,j-1)\} \\ DTW(x,y) = \min\{r(n,m)\} \end{cases} \tag{7}$$

DTW allows comparisons between time series of different lengths and identifies similar shapes, even if they are out of phase. However, it is computationally intensive, making it less practical for large datasets.

o   Longest Common Subsequence (LCSS) focuses on the longest matching subsequences between two time series while ignoring noise and distortions. For two sequences $S_x$ and $S_y$ of lengths n and m, the similarity is defined as:

$$M(i,j) = \begin{cases} 0 & ; i = 0 \text{ or } j = 0 \\ 1 + M(i-1, j-1) & ; x_i = y_j, \ i \geq 1 \text{ or } j \geq 1 \\ Max \begin{cases} M(i-1,j) \\ M(i,j-1) \end{cases} & ; x_i \neq y_j, \ i \geq 1 \text{ or } j \geq 1 \end{cases} \quad (8)$$

Where M(n,m) is calculated recursively:

$$M(i,j) = \begin{cases} 0 & ; i = 0 \text{ or } j = 0 \\ 1 + M(i-1, j-1) & ; (x_i - y_j) \leq \varepsilon, \ i \geq 1 \text{ or } j \geq 1 \\ Max \begin{cases} M(i-1,j) \\ M(i,j-1) \end{cases} & ; (x_i - y_j) > \varepsilon, \ i \geq 1 \text{ or } j \geq 1 \end{cases} \quad (9)$$

LCSS is robust to noise and suitable for comparing time series with different lengths. However, it heavily depends on the similarity threshold, which impacts its accuracy.

### 2-4-3. Geometric distances

Geometric distances focus on the spatial characteristics of trajectories, particularly their shapes. Examples include Hausdorff Distance, Discrete Frechet Distance, and SSPD (Symmetric Segment Path Distance).

o   The Hausdorff Distance measures the maximum mismatch between two trajectories, defined as:

$$Haus(X, Y) = Max\{\sup_{x \in X} \inf_{y \in Y} \|xy\|_2, \sup_{x \in X} \inf_{y \in Y} \|xy\|_2\} \quad (10)$$

o   Frechet Distance measures the similarity between curves by calculating the minimal "leash length" required to connect a dog and its owner walking along two separate paths. It is mathematically defined as:

$$D_{Frechet}(T^1, T^2) = \min_{w} \{\max_{k \in [0 \dots |w|]} \|w_k\|_2\} \quad (11)$$

Table 1 comprehensively evaluates the advantages and limitations of each paradigm.

Table1. Similarity methods

| Method | Advantages | Disadvantages | Category |
|---|---|---|---|
| **Euclidean Distance** | - Most straightforward and widely used criterion<br>- No need for parameter estimation | - Does not support local time shifts<br>- Inefficient with high-dimensional time series<br>- Sensitive to small changes in the time axis | Step-by-step measures |
| **DTW (Dynamic Time Warping)** | - Supports local scaling and order preservation of time series<br>- Handles time series of different | - Time-consuming<br>- Sensitive to noise<br>- High computational load | Elastic measures |

|  | lengths<br>- Measures local time shifts | - Incorrect clustering due to outliers<br>- Requires pairing all elements<br>- Not metric |  |
|---|---|---|---|
| **LCSS (Longest Common Subsequence)** | - Robust against noise<br>- Focuses on similar parts for clustering<br>- No need for data normalization | - Strongly depends on similarity threshold<br>- Zero-and-one similarity approach can lead to poor results<br>- Not metric and does not obey the triangle inequality | Elastic measures |
| **Hausdorff Distance** | - Measures spatial similarity between two routes<br>- Considers the farthest point in a set | - Not suitable for trends<br>- Complex calculations due to considering all points<br>- Limited in comparing paths<br>- Returns the maximum distance, ignoring overall similarity | Geometric measures |
| **Discrete Frechet Distance** | - Considers order and continuity of points<br>- Reduces complexity with discrete regression models | - Limited applicability to path comparison<br>- Maximum distance can overshadow detailed differences | Geometric measures |

## 4. Result

The finance dataset has been used to conduct prediction models. To optimize model performance, the Ridge parameter α of the robust Rank Ridge Regression model was determined using RidgeCV to minimize the mean squared error (MSE). Figure 1 demonstrates the optimal value of this parameter.

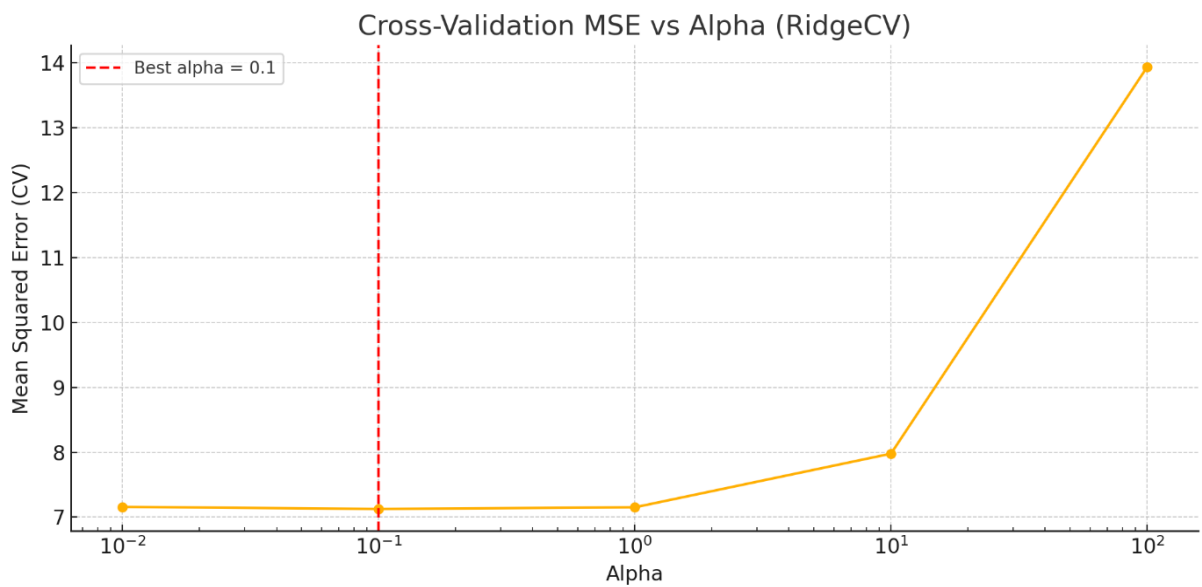

The Ridge Regression model was built to forecast Apple's daily closing price. To improve model accuracy, the past three-time lags[1] were created for the closing prices of both Apple and the predictors. The dataset was split into training and test sets, and the features were standardized to improve model performance. Model accuracy was evaluated using the Mean Squared Error (MSE) and the R² score, as shown in Table 1. The R² and MAE values for both the training and test datasets indicate the model's strong performance. The high R² value for test suggests that the model generalizes well and does not suffer from overfitting.

---

[1] Expanding window cross-validation was used to evaluate performance across various maximum lag values.

|  | RIDGE REGRESSION WITH LAGS |
|---|---|
| R² (TRAIN) | 0.992304404 |
| R² (TEST) | 0.960389253 |
| MAE (TRAIN) | 1.91299715 |
| MAE (TEST) | 3.263312581 |
| RMSE (TRAIN) | 2.548285132 |
| RMSE (TEST) | 4.313346635 |

Based on the Ridge Regression model, Figure 2 presents the prediction results for the AAPL test dataset. The blue line represents the actual test values, while the orange line indicates the predicted values. Considering that Ridge Regression is appropriate for data affected by multicollinearity, the accurate predictions on the test set—illustrated in Figure 2—demonstrate the model's accuracy.

Figure 2. Ridge Regression with Lag Features

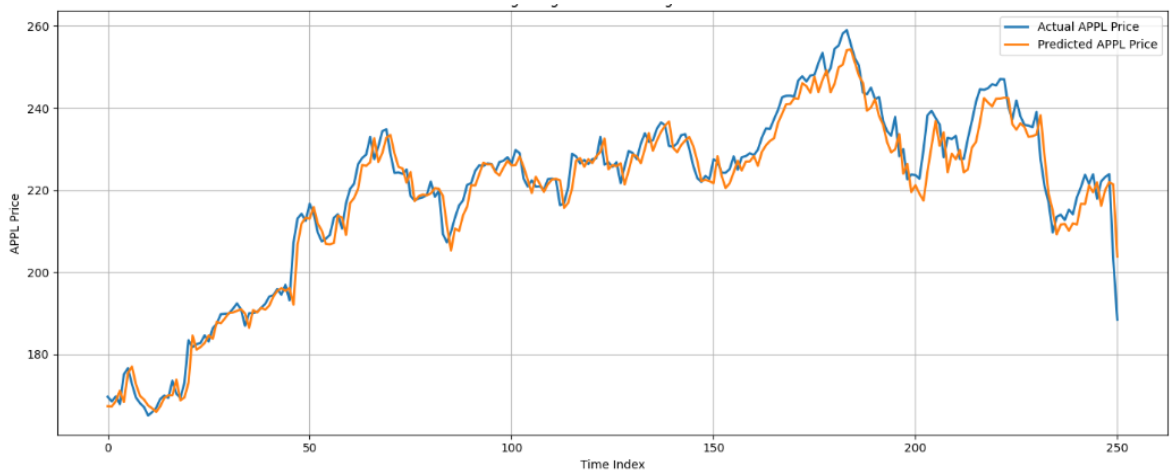

**Confidence Intervals for Model Predictions**

Coverage and width are the two primary properties that define the effectiveness of confidence intervals. This study computes the model's confidence intervals using standard deviation and time series similarity methods to assess which technique better satisfies the desired properties of accurate confidence intervals.

Traditional confidence intervals are derived based on the standard deviation, assuming a 95% confidence level. Based on the dataset used in this study, the coverage of the conventional confidence interval is 95.22%, meaning that 95.22% of the time, the actual value falls within the interval. Another important property for evaluating the effectiveness of a confidence interval is the Mean CI Width; a narrower width indicates more precise intervals. The mean CI width for our dataset is 14.57.

Figure 3. conventional confidence interval with standard deviation

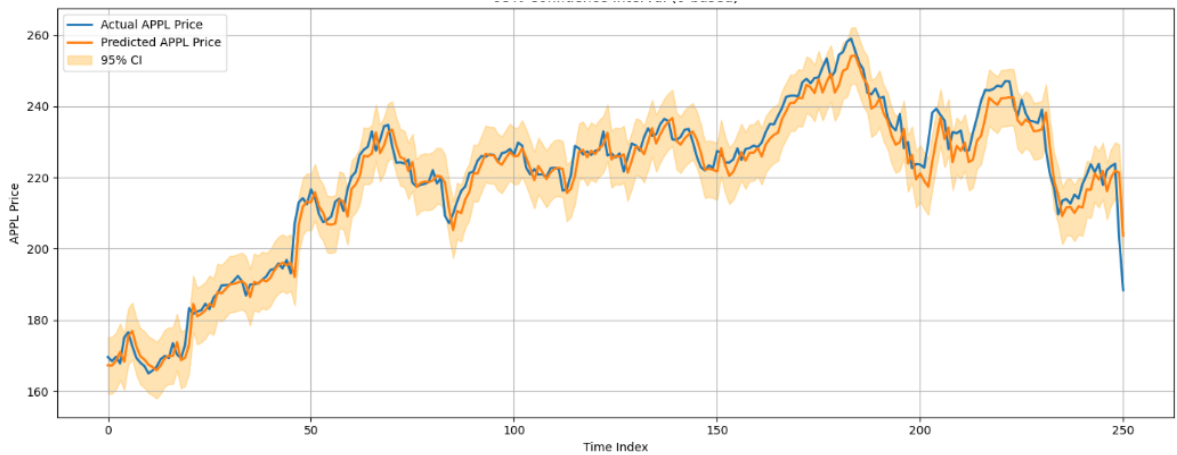

The main question of this research is: *Are time series similarity methods effective enough to construct reliable intervals comparable to those based on standard deviation?*. To evaluate this, confidence intervals were constructed using various time series similarity methods, including Dynamic Time Warping (DTW), Longest Common Subsequence (LCSS), Hausdorff distance, Time Warp Edit Distance (TWED), and Fréchet distance. Figure 4 illustrates the intervals generated by each method, highlighting the extent to which the actual values are covered within these intervals.

Figure4. confidence interval with time series similarity methods.

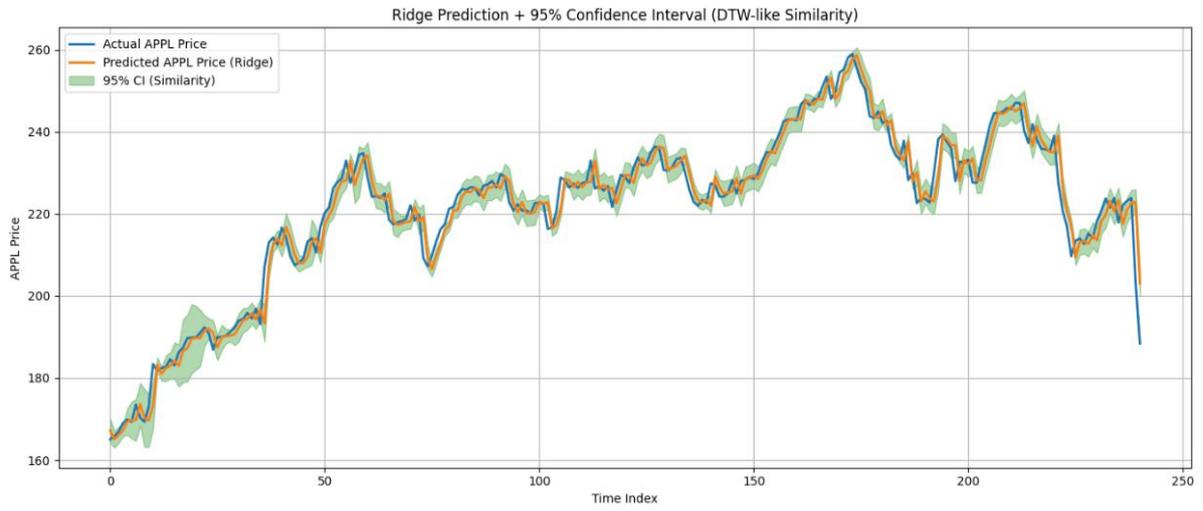

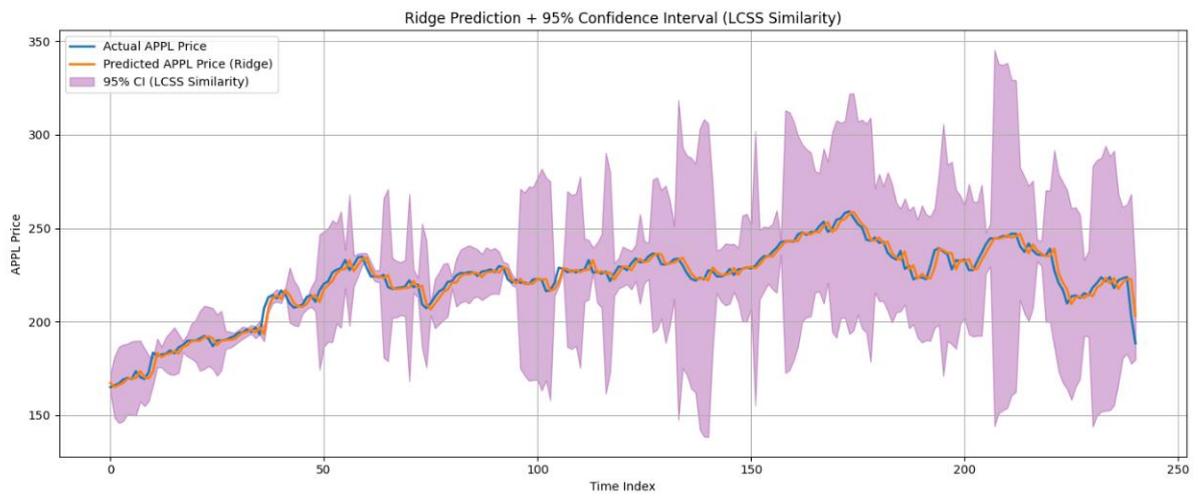

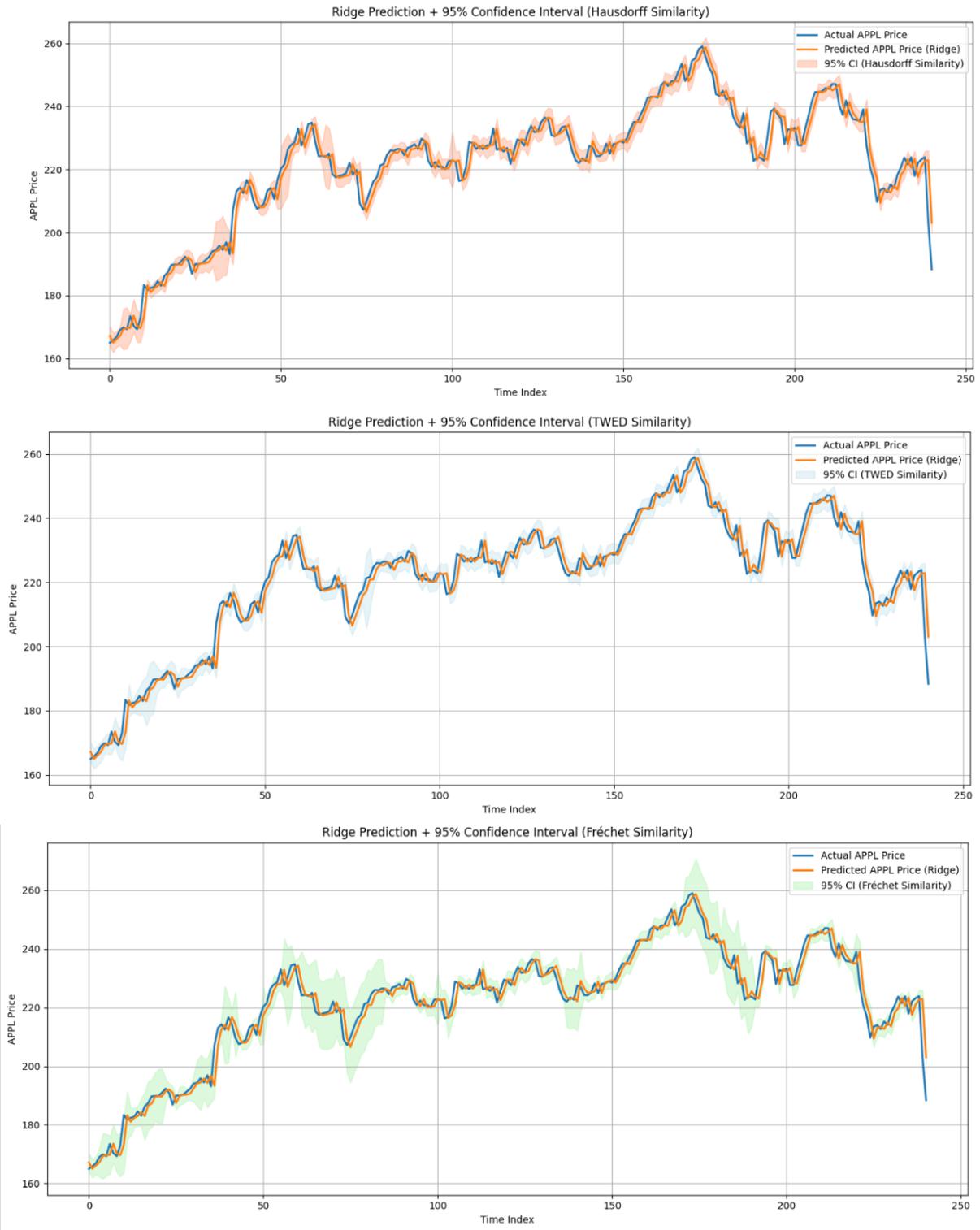

Table 2 summarizes the coverage and mean confidence interval width associated with each method.

Table2. Evaluation of CI Coverage and Width Using Conventional and Time Series Similarity Approaches

| METHOD | CI COVERAGE (%) | MEAN CI WIDTH |
|---|---|---|
| CONVENTIOAL | 95.22 | 14.57 |
| **DTW-LIKE** | 63.07 | 5.86 |
| **LCSS** | 94.61 | 59.34 |
| **HAUSDORFF** | 63.49 | 6.22 |
| **TWED** | 66.8 | 6.48 |
| **FRÉCHET** | 75.1 | 10.68 |

Figure 5. Comparison of Confidence Interval Coverage and Width Across Conventional and Time Series Similarity

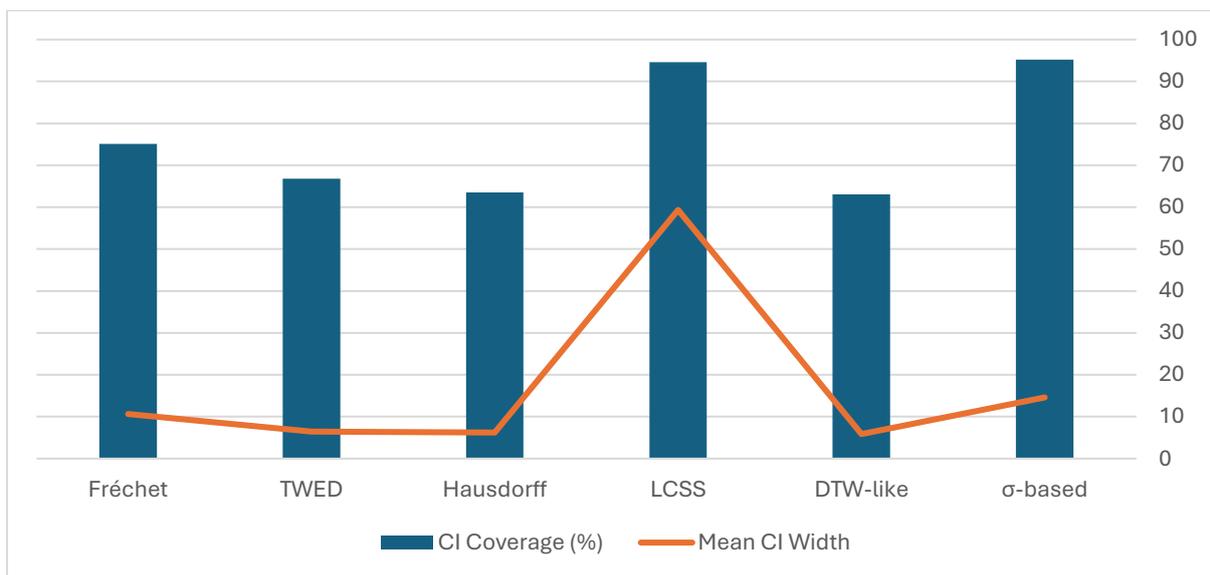

## 5. Discussion

Table 2 highlights the trade-off between reliability (measured by coverage) and precision (measured by mean confidence interval width) across six different methods for constructing confidence intervals (CIs). This study compares the performance of these methods in capturing true values within the estimated intervals.

The first method, a σ-based approach (commonly referred to as the conventional confidence interval), demonstrates strong performance with a coverage rate of 95.22% and a moderate mean width of 14.57. Among all the methods evaluated, this approach yields the highest coverage, indicating that it reliably includes the actual values within the interval. However, while reliable, its precision is only moderate compared to some of the other methods.

The LCSS (Longest Common Subsequence) method also achieves high coverage at 94.61%, closely matching the conventional method. However, it produces significantly wider intervals (mean width of 59.34), suggesting that while it captures true values effectively, it lacks precision and may result in overly conservative estimates.

In contrast, the DTW-like, Hausdorff, and TWED methods generate much narrower confidence intervals (with mean widths of 5.86, 6.22, and 6.48, respectively), which indicates greater precision. However, this precision comes at the cost of lower coverage—ranging from 63.07% to 66.8%—which implies that these methods may fail to capture the actual values as consistently as the σ-based or LCSS methods.

The Fréchet distance-based method offers a balance between coverage and precision, with a coverage of 75.1% and a mean width of 10.68. While not as reliable as the σ-based or LCSS approaches, it outperforms the other similarity-based methods in terms of maintaining a more favorable balance.

Overall, the results illustrate a clear trade-off: methods that produce narrower (more precise) intervals tend to have lower coverage, while those with higher coverage often result in wider (less precise) intervals. The choice of method should therefore depend on the specific application and whether reliability or precision is prioritized.

6. Conclusion

In the realm of statistical modeling, two primary objectives—prediction and explanation—guide the analytical process. When forecasting is the focus, it becomes essential to consider the uncertainties that arise when estimating unknown outcomes. Historically, confidence intervals created from standard deviations have provided a structured method for quantifying this uncertainty and enabling an assessment of how closely predicted values align with their actual counterparts. This traditional approach implicitly seeks to embody the behavioral similarities between observed and predicted data points. Nevertheless, recent advancements in similarity-based methodologies offer innovative alternatives to conventional variance-focused techniques, especially in settings characterized by extensive datasets or a significant number of explanatory variables. This article seeks to explore methods that can effectively reduce uncertainty in confidence interval estimation. By investigating both traditional and similarity-based approaches, the goal is to identify which of these methods can yield tighter confidence intervals under similar conditions, ultimately leading to greater precision and more informative results. Addressing the challenge of uncertainty is paramount, as it underpins the reliability of predictions and enhances decision-making processes in various applications.

In conclusion, this study illustrates a significant trade-off between reliability and precision in confidence interval construction methods for capturing true values. The evaluation of six distinct methods highlights the differential performance in terms of coverage rates and mean confidence interval widths. The conventional σ-based approach emerged as the front-runner with the highest coverage rate (95.22%), ensuring that true values are reliably included within the intervals, albeit with moderate precision due to its wider interval width. The LCSS method closely follows, achieving high coverage at 94.61%, but its considerably wider intervals indicate a trade-off in precision, resulting in overly conservative estimations.

On the other hand, methods such as DTW, Hausdorff, and TWED demonstrate exceptional precision with notably narrower intervals. However, this precision is accompanied by lower coverage rates (ranging from 63.07% to 66.8%), suggesting a propensity to miss some actual values. The Fréchet distance-based method provides a more balanced approach, achieving reasonable coverage (75.1%) while maintaining a narrower mean confidence interval width.

Overall, the results underscore the necessity for careful consideration of the chosen method based on the specific context of the analysis. Researchers and practitioners must weigh their priorities—be it a higher reliability or greater precision—in order to make informed decisions that align with their forecasting objectives. This article emphasizes the importance of tailoring confidence interval estimation to meet the demands of various applications, recognizing that no one-size-fits-all solution exists in the pursuit of reducing uncertainty in predictions.

For future studies, it would be beneficial to explore more advanced similarity-based methods that could further enhance the accuracy of confidence intervals in predicting outcomes. Specifically, researchers could investigate the application of state-of-the-art techniques such as deep learning-based similarity measures, which have demonstrated strong performance across various domains. Additionally, methods incorporating dynamic time warping (DTW) and advanced metric learning algorithms could be employed to capture complex patterns and relationships in high-dimensional data more effectively.